\documentclass[amsmath,floatfix,aps,twocolumn]{revtex4}

\usepackage{amssymb}
\usepackage{graphicx}
\usepackage{hyperref}
\usepackage{epstopdf}

\newcommand{\torate}[1]{\overset{#1}{\to}}
\newcommand{\rt}{\lbrace\mathbf{r}(t)\rbrace}

\begin{document} \date{\today}
\title{Optimization of collective enzyme activity via spatial localization}

\author{Alexander Buchner}
\author{Filipe Tostevin}
\author{Florian Hinzpeter}
\author{Ulrich Gerland}
\email{gerland@lmu.de}
\affiliation{Arnold Sommerfeld Center for Theoretical Physics and Center for
NanoScience, Ludwig-Maximilians-Universit\"at, 80333 M\"unchen, Germany}

\begin{abstract} 
The spatial organization of enzymes often plays a crucial role in
the functionality and efficiency of enzymatic pathways. To fully understand
the design and operation of enzymatic pathways, it is therefore crucial to
understand how the relative arrangement of enzymes affects pathway function. 
Here we investigate the effect of enzyme localization on the flux of a minimal 
two-enzyme pathway within a reaction-diffusion model. We consider different 
reaction kinetics, spatial dimensions, and loss mechanisms for
intermediate substrate molecules. Our systematic analysis of the different
regimes of this model reveals both universal features and distinct
characteristics in the phenomenology of these different systems. In particular,
the distribution of the second pathway enzyme that maximizes the reaction flux
undergoes a generic transition from co-localization with the first enzyme when
the catalytic efficiency of the second enzyme is low, to an extended profile
when the catalytic efficiency is high. However, the critical transition point
and the shape of the extended optimal profile is significantly affected by
specific features of the model. We explain the behavior of these different
systems in terms of the underlying stochastic reaction and diffusion processes
of single substrate molecules.
\end{abstract} 
\maketitle

\clearpage
\section{Introduction}

The action of enzymes is essential for nearly all processes in living cells.
Often these enzymes are organized into large multi-molecular complexes
associated with specific functional tasks \citep{Srere1987}, and this
organization can be crucial to the successful operation of the enzymatic system.
These ``molecular factory'' assemblies, in which the product of one enzymatic
reaction becomes the substrate for the next, are common in metabolic pathways of
both prokaryotes and eukaryotes. Examples include the cellulosome
\citep{Bayer1998}, the pyruvate dehydrogenase complex \citep{deKok1998} and
glycolytic enzymes \citep{Campanella05}.  In some cases, such as the cellulosome
\citep{Bayer1998}, enzymes are arranged on an inert scaffold in a specific way.
In others, such as tryptophan synthase complexes \citep{Dunn2008}, direct
enzyme-enzyme interactions lead to self-assembly into a complex. 

Despite the ubiquity of these multi-enzyme complexes, we still lack a deep
understanding of the consequences of particular arrangements of enzymes for
metabolic pathway operation. Many advantages of co-localization have been
proposed \citep{Gaertner1978, Ovadi1991, Schuster1991}, particularly via the
direct transfer or ``channeling'' of substrates from one enzyme to another. For
example, reducing the transit time of pathway intermediates between enzymes can
minimize the loss of unstable intermediates or the interference of competing
pathways. Channeling could also potentially enhance the local density of
substrates in the vicinity of the enzymes and reduce exposure to 
toxic intermediates; however, whether or not these effects can actually occur has been
disputed \citep{Cornish1991, Mendes1992, Cornish1993, Mendes1996}. On the other
hand, compartmentalization of metabolic enzymes can also increase the flux of
biosynthetic pathways \cite{Conrado2007}, indicating that the pathway kinetics
can be influenced by localization even in the absence of direct channeling.
Similar questions about the role of co-localization also arise in the context of
protein signaling cascades. For example, there the clustering of enzymes can
generate a greater amplification of the signal than distributing enzymes
\citep{vanAlbada2007, Mugler2012}. However, differing functional criteria between
signaling scenarios, where discrimination between different inputs is crucial,
and metabolic systems, where maintaining a specific flux may be more desirable,
mean that these systems are likely subject to different design pressures. More
generally, little is known about the effects of the placement of enzymes beyond
simple co-localization or clustering scenarios.

Recently there has been a growing focus on the experimental study of
colocalized enzymes. Techniques have been developed that
allow for the attachment of enzymes to a scaffold \citep{Conrado2008}, which was
shown to significantly increase the yield of the mevalonate production pathway
\citep{Dueber}. The ``single-molecule cut-and-paste'' technique \citep{Kufer08}
allows for the positioning of enzymes on a surface with nanometer precision. DNA
origami permits the highly-controlled production of three-dimensional structures
\citep{Douglas2009}, enabling the quantitative study of the effects of more
complex spatial arrangements of enzymes. Over the last few years much progress
has been made in engineering of artificial enzymatic pathways on DNA
\citep{Niemeyer2002, Wilner2009, Idan2013} and RNA assemblies
\citep{Delebecque2011}, even {\em in vivo}. In particular, a distance-dependence
of the activity of a pathway consisting of glucose oxidase (GOx) and the
horseradish peroxidase (HRP) was demonstrated \citep{Yan2012}: when the enzymes
are brought closer together, the efficiency of the two enzyme complex increases.

Here we study theoretically the impact of enzyme positioning on the flux of
pathways. It has been demonstrated previously \citep{US} that in a simple
linear reaction-diffusion model, in different parameter regimes co-localization can
increase or decrease the pathway flux compared to the uniform distribution of
enzymes. In this paper, we extend these results to a range of reaction-diffusion
systems. In particular, we consider also nonlinear reactions and different spatial 
dimensions. We demonstrate that the qualitative features of these diverse models
are similar. In general, a transition occurs as a function of the effective
reaction rate between regimes in which clustering or distributing enzymes in
space generates a higher pathway efficiency. We calculate the optimal enzyme
distribution that maximizes the efficiency of the pathway. The universal nature
of our results in these diverse systems shows that the observed transitions
arise from general properties of reactions and diffusion, and highlights the
applicability of the observed behavior to diverse biochemical pathways.

\section{Model} \label{sec:Model}

\begin{figure}
	\includegraphics{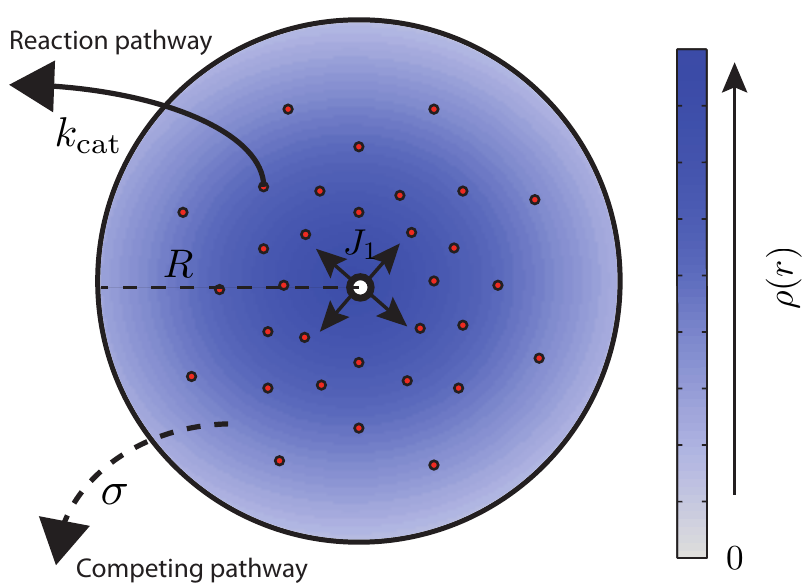}
	\caption{Illustration of the type of reaction-diffusion systems considered in
		this paper. A cluster of $E_1$ enzymes at the center of the system acts as a
		source of intermediate $I$ with total production rate $J_1$.  Intermediates
		diffuse from this center, and can either be converted to the desirable
		product by the enzymes $E_2$ (red circles), via a Michaelis-Menten reaction
		with catalytic rate $k_{\rm cat}$, or can be lost to a competing pathway
		with the spatially-uniform rate $\sigma$.  We coarse-grain the positions of
		$E_2$ in space into the continuous distribution $e(\mathbf{r})$, and the
		local density of intermediates as $\rho(\mathbf{r})$. While a
		two-dimensional system with a hard-wall outer boundary is shown here for
		clarity, we consider systems of all dimension as well as different
		outer boundary conditions.}
	\label{fig:Model}
\end{figure}

We consider a simple model reaction pathway consisting of two enzymatic reaction
steps. In the first reaction step, an enzyme $E_1$ converts a substrate $S$ into
an intermediate $I$; subsequently, a second enzyme $E_2$ converts $I$ into the
final product of the pathway, $P$. We are interested in how the spatial
organization of the enzymes affects the efficiency of the pathway $S
\torate{E_1} I \torate{E_2} P$ in converting substrate $S$ to product $P$.  To
this end we assume that the $E_1$ enzymes are fixed in position, and examine the
impact of the location of the $E_2$ enzymes relative to $E_1$. In this scenario,
$E_1$ enzymes act as a source of intermediate $I$, with a total production rate
$J_1$. In order to additionally include possible undesirable non-specific
competition for the intermediate by secondary pathways, or decay in the case
that $I$ is unstable, we also allow for the conversion of $I$ into an
alternative waste product $Q$. Under the assumption that these processes are
independent of the spatial arrangement of $E_2$ enzymes, they are simply modeled
as a first-order reaction with a constant, position-independent, rate $\sigma$.
The density of intermediate $I$, $\rho({\mathbf{r}},t)$ can then be modeled by
the reaction-diffusion equation
\begin{equation} \label{3d_reaction_diffusion}
	\frac{\partial\rho(\mathbf{r},t)}{\partial t}=D\nabla^2\rho(\mathbf{r},t)-
	\frac{k_{\rm cat}e(\mathbf{r})\rho(\mathbf{r},t)}{K_M+\rho(\mathbf{r},t)}
	-\sigma\rho(\mathbf{r},t),
\end{equation}
where $D$ is the diffusion constant of $I$ and $e(\mathbf{r})$ is the (static)
density of $E_2$ enzymes. The model is illustrated in Fig.~\ref{fig:Model}. In writing
Eq.~\ref{3d_reaction_diffusion} we have assumed that the conversion of $I$ to
$P$ by the enzyme $E_2$ can be described by standard Michaelis-Menten kinetics
with catalytic rate $k_{\rm cat}$ and Michaelis constant $K_M$. We 
implement the production of intermediate by $E_1$ enzymes through boundary
conditions to Eq.~\ref{3d_reaction_diffusion}. In this work we restrict
ourselves to the case of a uniform source at the inner boundary, $-D
\nabla\rho(\mathbf{r}_{\rm in})\cdot\mathbf{n}(\mathbf{r}_{\rm in})= J_1/A_{\rm
in}$, where $\mathbf{n}(\mathbf{r}_{\rm in})$ is the unit vector normal to the
boundary and $A_{\rm in}$ is the area of the inner boundary; thus the total
influx integrated over the boundary equals the production rate $J_1$.  For the
outer boundary we limit ourselves to reflective
($\nabla\rho(\mathbf{r}_{\rm out})\cdot\mathbf{n}(\mathbf{r}_{\rm out})=0$) or
absorbing ($\rho(\mathbf{r}_{\rm out})=0$) boundary conditions. The former could
represent the confinement of the intermediate reaction product by the membranes
of a cell or organelle, while the latter might describe an intermediate that can
easily cross the membrane and be lost to the extracellular environment. We note 
that our treatment could easily be generalized to mixed boundary conditions 
representing partial confinement. 

In the following we will be concerned only with the steady-state flux through
the reaction pathway. At steady-state form, Eq.~\ref{3d_reaction_diffusion} can
be recast into the dimensionless form
\begin{equation} \label{general_dimensionless}
	0=\nabla^2\rho'(\mathbf{r'})
		-\frac{\alpha e'(\mathbf{r'})\rho'(\mathbf{r'})}{1+\gamma\rho'(\mathbf{r'})}
		-\beta\rho'(\mathbf{r'}),
\end{equation}
where $\mathbf{r'}$ denotes that the spatial coordinate has been rescaled by a
characteristic length-scale of the system, $R$, which we will take to be the
system size; and $\rho'$ indicates that the density has further been rescaled
such that the total production rate of intermediate is equal to $1$.
Additionally, we have defined the rescaled enzyme density
$e'(\mathbf{r})=e(\mathbf{r})/\bar{e}$ with $\bar{e}=V^{-1}\int_V
e(\mathbf{r}){\rm d}\mathbf{r}$ the average enzyme density over the system
volume $V$. The dimensionless parameters $\alpha=(k_{\rm
cat}\bar{e}/K_M)(R^2/D)$ and $\beta=\sigma(R^2/D)$ respectively capture the
relative timescales of reactions with $E_2$ and with secondary pathway enzymes,
compared to the typical time to diffuse a distance $R$. The parameter
$\gamma=J_1R^{2-d}/(K_MD)$, where $d$ is the spatial dimension, represents the
rate of influx of intermediate relative to the level at which $E_2$ enzymes
become saturated, and includes the effect of varying the activity of $E_1$
enzymes via the intermediate production rate $J_1$. In the following we drop the
prime notation and work exclusively with the dimensionless system; this should
not be a source of confusion.  

Integrating Eq.~\ref{general_dimensionless} and applying the boundary
conditions leads to the flux-conservation equation
\begin{equation} \label{conservation_equation}
	1=
	\underbrace{
		\int_V\frac{\alpha e(\mathbf{r})\rho(\mathbf{r})}{1+\gamma\rho(\mathbf{r})}
			{\rm d}\mathbf{r}
	}_{\displaystyle J_2/J_1}
	+\underbrace{
		\int_V\beta\rho(\mathbf{r}){\rm d}\mathbf{r}-
		\int_{\partial V}\nabla\rho(\mathbf{r})\cdot\mathbf{n}(\mathbf{r})
			{\rm d}\mathbf{r}
	}_{\displaystyle J_{\rm loss}/J_1}.
\end{equation}
On the left hand side we have the (rescaled) production of intermediate by
$E_1$. This must be balanced by the flux of reactions by $E_2$ enzymes,
$J_2/J_1$, plus the loss of intermediate. This loss can occur to secondary
pathways (the second term on the right hand side of
Eq.~\ref{general_dimensionless}), and via escape at the boundaries of the system
(the third term, where $\partial V$ is the outer boundary of the system that is
not a source of intermediate). Assuming that the efficiency of the conversion of
substrate to intermediate by $E_1$ is independent of the localization of $E_2$
enzymes, such that $J_1$ is constant, the efficiency of the system can be
described by the ratio $J_2/J_1$, the fraction of intermediates that are
converted into the correct product $P$. In this work we will examine how
changing $e(\mathbf{r})$ affects the pathway efficiency $J_2/J_1$. To compare
different enzyme profiles on an equal footing, the total amount of $E_2$ is held
constant via the condition $V^{-1}\int_Ve(\mathbf{r}){\rm d}\mathbf{r}=1$.

In the remainder of this paper we systematically characterize the effects of
varying $E_2$ localization in different regimes of
Eq.~\ref{general_dimensionless}. In the section `Linear reaction models' we
will focus on the low density limit of the intermediate product, in which the
rate of reaction with $E_2$ becomes linear in $\rho(\mathbf{r})$. It was
previously shown \citep{US} that in an open one-dimensional system where
intermediate is lost at an absorbing boundary, different parameter regimes exist
in which the optimal enzyme profile consists of either co-localization of $E_2$
with $E_1$, or a configuration wherein only a fraction of $E_2$ enzymes are
co-localized and the remainder are distributed over a finite region. Here we
extend these results to consider closed systems where the loss of intermediate
occurs only via position-independent secondary reactions, and to three- and
two-dimensional systems. Finally, in the section `Non-linear reactions' we
consider also the full non-linear reaction model. In all cases we observe a
transition in the optimal profile from co-localized to distributed as a function
of the system parameters, analogous to that reported in the specific minimal
model of \citet{US}, demonstrating the generality of the underlying physics.
However, we also highlight qualitative differences in the phenomenology of these
different regimes.

\section{Results}
\subsection{Linear reaction models} \label{sec:linear}
\subsubsection{Enzyme exposure}

\begin{figure*}
	\includegraphics{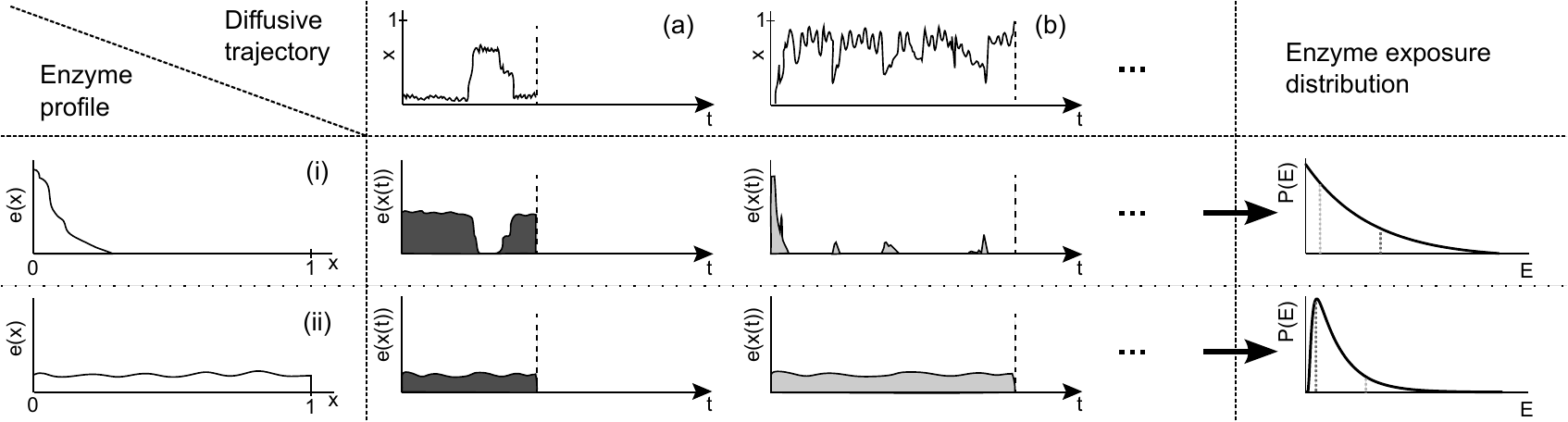}
	\caption{A schematic illustration of the underlying concept  of ``enzyme
		exposure''. 
		Diffusion generates an ensemble of trajectories of intermediate molecules
		(top row), which have different enzyme exposure values in the presence of
		different enzyme distribution patterns (left column). For example, the
		diffusive path (a) spends a relatively long time in the vicinity of the
		origin. This leads to a higher enzyme exposure value $E=\int_0^{t_{\rm
		escape}} e(x(t))dt$ (shaded areas) for an enzyme distribution that is
		clustered near the origin [distribution (i), middle row]. Trajectory (b),
		which spends little time near the origin, leads to a very small exposure
		value in the presence of distribution (i). For a more uniformly distributed
		profile [distribution (ii), bottom row] the exposure value is determined
		primarily by how long the particle stays in the system. Enumerating the
		value of $E$ for all possible diffusive trajectories leads to the
		$e(x)$-dependent distribution of enzyme exposures $P(E)$ (right column).
	}
	\label{fig:Ecartoon}
\end{figure*}

To understand the impact of different enzyme configurations on the overall
pathway flux, the concept of integrated ``enzyme exposure'' has proven to 
be useful \citep{US}. It allows for the decomposition of the reaction flux of linear
systems into two factors, one that depends only on the enzyme distribution
$e(\mathbf{r})$ and describes the diffusive dynamics of the system, and another
that is independent of $e(\mathbf{r})$ but captures the reaction dynamics. 
The concept is best explained with the help of a thought experiment where we 
first consider the dynamics of individual intermediate molecules in the absence 
of any $E_2$ enzymes, also shown schematically in Fig.~\ref{fig:Ecartoon}. We
introduce a single intermediate molecule at $t=0$ at the source, and track its
stochastic path until it leaves the system, either through the boundary of the
system or via a reaction with a competing pathway. We denote the time at which
the trajectory ends, either by escaping at the system boundary or through a
competing reaction, as $t_{\rm escape}$. By repeatedly applying this procedure
for many such molecules, we can generate an ensemble of trajectories
$\mathbf{r}(t)$ through the system that is independent of the distribution of
$E_2$ enzymes. 

\begin{figure}
\includegraphics{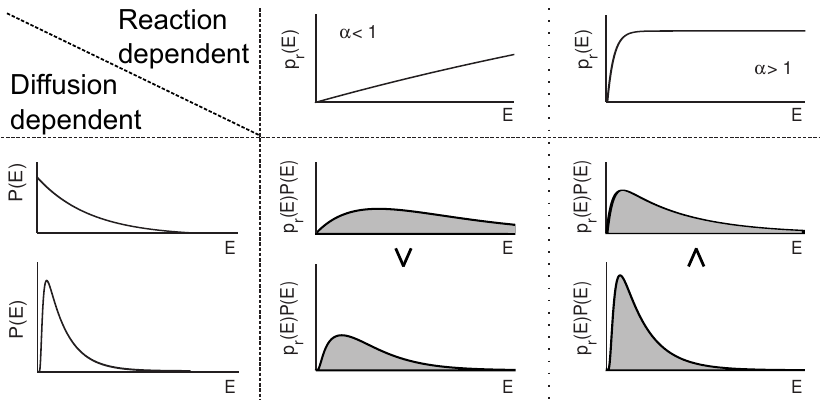}
	\caption{The reaction efficiency can be calculated as the overlap integral of the
$e(\mathbf{r})$-dependent enzyme exposure distribution $P(E)$, and the $\alpha$-dependent
reaction probability $p_r(E)$, according to $J_2/J_1=\int_0^\infty
P(E)p_r(E){\rm d}E$. Broader exposure distributions, which maximize the
likelihood of large-$E$ trajectories, are preferable when $\alpha$ is small
(middle column).  Narrower $P(E)$ distributions, which minimize the likelihood
of small values of $E$, are favored when $\alpha$ is large (right).}
	\label{fig:reactioncartoon}
\end{figure}

Next, we suppose that we were to re-introduce $E_2$ enzymes according to the
distribution $e(\mathbf{r})$. For each of the diffusive intermediate 
trajectories generated above, the instantaneous propensity of reaction with an
$E_2$ enzyme is given by $\alpha e(\mathbf{r}(t))$ (in the linear reaction
regime). For each trajectory, the survival probability $S(t)$ that no reaction 
has occurred up to the time $t$ follows the differential equation
$\dot{S}(t)=-\alpha e(\mathbf{r}(t))S(t)$. We can therefore straightforwardly 
calculate the probability that a reaction would have occurred at some point
along the trajectory as $1-\exp\left[-\alpha\int_0^{t_{\rm escape}}
e(\mathbf{r}(t)){\rm d}t\right]$. 

Finally, we define the enzyme exposure for each individual trajectory to be 
\begin{equation} \label{eq:E_def}
	E=\int_0^{t_{\rm escape}}e(\mathbf{r}(t)){\rm d}t.
\end{equation}
The ensemble of possible trajectories in the system $\mathbf{r}(t)$, each with a
characteristic $t_{\rm escape}$, therefore generates a distribution of enzyme
exposure values, $P(E)$. This distribution is a function of the arrangement of
$E_2$ enzymes $e(\mathbf{r})$ via Eq.~\ref{eq:E_def}, but importantly is
independent of the reaction with $E_2$ enzymes, since the diffusive trajectories
were generated in the absence of such reactions. The probability of reaction
along a trajectory is then given by $p_r(E)=1-\exp[-\alpha E]$, which depends on
the reaction parameter $\alpha$ but crucially not on the $E_2$ distribution
itself. The overall probability of reaction is recovered by the expression 
\begin{equation} \label{eq:efficiency_traj}
	\frac{J_2}{J_1}=\int_0^\infty P(E)p_r(E){\rm d}E,
\end{equation} 
which ensures a proper weighting of the likelihood of a particular trajectory
occurring (see the Supplementary Material \cite{SI} for a derivation showing the
equivalence of Eq.~\ref{eq:efficiency_traj} with the expression for $J_2/J_1$
defined in Eq.~\ref{conservation_equation}). Thus, as depicted schematically in
Fig.~\ref{fig:reactioncartoon}, we have decomposed the reaction-diffusion
dynamics of Eq.~\ref{general_dimensionless} into a diffusion- and
$e(\mathbf{r})$-dependent component $P(E)$, and a reaction-dependent component
$p_r(E)$, with the efficiency of the reaction pathway determined by the product
of these two distributions. 

Interestingly, using Eq.~\ref{conservation_equation} the reaction efficiency can
be rewritten as 
\begin{equation}
	\frac{J_2}{J_1}=1-\int_0^{\infty}P(E)e^{-\alpha E}{\rm d}E
	=1-\frac{J_{\rm loss}}{J_1},
	  \label{Inverse_laplace}
\end{equation}
wherein we see that the fraction of intermediate molecules lost via the reaction
$I\to Q$ or through the boundary, $J_{\rm loss}/J_1$, takes the form of the
Laplace transform of $P(E)$, with transform variable $\alpha$. It is generally
more straightforward to calculate $J_{\rm loss}$ as a function of $\alpha$ for a
given $E_2$ profile $e(\mathbf{r})$ and to compute $P(E)$ by performing an
inverse Laplace transformation, than it is to calculate $P(E)$ directly from
considering individual diffusive trajectories.

\subsubsection{A competing pathway} \label{leakage}

We now consider the case where intermediates are unable to cross a cell membrane
and therefore cannot escape via the boundaries of the system, but can be lost to
a competing, spatially-uniform, reaction pathway. That is, all intermediate
molecules ultimately end up as either the desirable product $P$ or undesirable
product $Q$. For a one-dimensional system in the linear, low-density, regime of 
the Michaelis-Menten reaction, we have the reaction-diffusion equation 
\begin{equation}\label{reaction_diffuison_leakage}
	0=\partial^2_x\rho(x)-\alpha e(x)\rho(x)-\beta\rho(x)\;
\end{equation}
with a source boundary condition at the left edge,
$\partial_x\rho\vert_{x=0}=-1$, and a reflecting boundary at the right edge,
$\partial_x\rho\vert_{x=1}=0$.  The parameters $\alpha$ and $\beta$, defined
above, reflect the relative reactivities of intermediate with $E_2$ and 
competing pathway enzymes respectively, in units of the typical time to diffuse
a distance of the system size $R$. Consequently, $\alpha$ and $\beta$ can also
be interpreted as describing the system size in units of the typical distance
from the source at $x=0$ that an intermediate molecule will diffuse before
leaving the system via reaction with $E_2$ and via the competing pathway,
respectively.  When $\beta\ll1$, intermediate molecules will typically be able
to explore the entire system, and therefore we should expect that the spatial
arrangement of enzymes will have little effect on the reaction flux, since the
intermediate will be exposed to each enzyme regardless of where it is placed. In
contrast, when $\beta\gg1$ very few intermediate molecules will diffuse far from
the source and we should expect that the amount of enzyme located close to the
source will have a strong influence on the pathway efficiency.

We begin by examining the case where $E_2$ enzymes have the uniform 
density $e_{\rm u}(x)=1$ throughout the domain $x\in[0,1]$. Substituting into
Eq.~\ref{reaction_diffuison_leakage} leads to the straightforward solution
$\rho_{\rm u}(x)=\cosh\left[(1-x)\sqrt{\alpha+\beta}\right]/
(\sqrt{\alpha+\beta}\sinh\sqrt{\alpha+\beta})$. From this expression the
reaction efficiency can be calculated using the definitions in
Eq.\ref{conservation_equation}, and is given by
\begin{equation}
	\left(\frac{J_2}{J_1}\right)_{\rm u}=\frac{\alpha}{\alpha+\beta}.
\end{equation}
Next we suppose that all $E_2$ enzymes are clustered at $x=0$, co-localized with
$E_1$. This is represented by the distribution $e_{\rm c}(x)=\delta(x)$, which 
leads to the solution $\rho_{\rm c}(x)=\rho_0
	\cosh\left[(1-x)\sqrt{\beta}\right]/\cosh\sqrt{\beta}$, where $\rho_0$ can be
found by imposing the flux conservation equation
(\ref{conservation_equation}) with $\gamma=0$. Ultimately, this leads to a
reaction efficiency of 
\begin{equation}
	\left(\frac{J_2}{J_1}\right)_{\rm c}=
		\frac{\alpha}{\alpha+\beta^{1/2}\tanh\beta^{1/2}}.
\end{equation}

Comparing these two expressions, we can see that for given values of $\alpha$
and $\beta$ the clustered $E_2$ configuration always generates a higher reaction
flux than a uniform distribution of $E_2$ since
$\tanh\beta^{1/2}\leq\beta^{1/2}$. (Note that this situation changes
considerably if we allow for escape of intermediate at the boundary in addition
to loss via secondary reactions. For full details, see Appendix.) Intuitively,
this is because secondary reactions, parametrized by $\beta$, limit how far
intermediate molecules diffuse away from the source at $x=0$.  Thus for a
uniform profile, the effective enzyme density that intermediate molecules will
experience is reduced; for large $\beta$, this reduction is by a factor of order
$\beta^{1/2}$. This effect can also be quantified by studying the enzyme
exposure distributions corresponding to the two enzyme profiles, which are
\cite{SI}
\begin{subequations}
	\begin{align}
	P_{\rm u}(E)& =\beta e^{-E\beta},\\
	P_{\rm c}(E)& =\beta^{1/2}\tanh\beta^{1/2}
		e^{-E\beta^{1/2}\tanh\beta^{1/2}}.
	\end{align}
\end{subequations}
We see that $P_{\rm u}(E)$ is more concentrated near $E=0$ than $P_{\rm c}(E)$;
thus a higher proportion of trajectories rapidly react via secondary pathways
before being exposed to a significant level of $E_2$ enzyme. These intermediate
molecules therefore have a low probability of reacting with $E_2$ leading to a
low pathway efficiency.

We now turn to the question of what is the $E_2$ profile that maximizes the
reaction efficiency. We investigated this by performing a numerical optimization
of $e(x)$ on a discrete lattice of $N$ sites, as described previously
\citep{US}.  Briefly, we use an evolutionary algorithm with mutation and mixing.
We begin each optimization step with a trial enzyme profile $e(x)$. We generate
50 mutations of this profile, by selecting one lattice site at random and moving
a random fraction of the $E_2$ enzymes at this site to another randomly-chosen
site. For each of these modified $e(x)$ profiles, the discrete
reaction-diffusion system (an order-$N$ linear system) is solved and $J_2/J_1$
calculated. As the initial profile for the next mutation round, we take the mean
of the ten most-efficient mutant configurations in the previous round. We have
found this procedure to achieve more rapid and robust convergence than a simple
Monte Carlo exploration of the space of possible configurations.  The optimal
profiles reported below are the configurations with the highest reaction
efficiency that occurred at any point during the optimization process.

\begin{figure*}
	\includegraphics{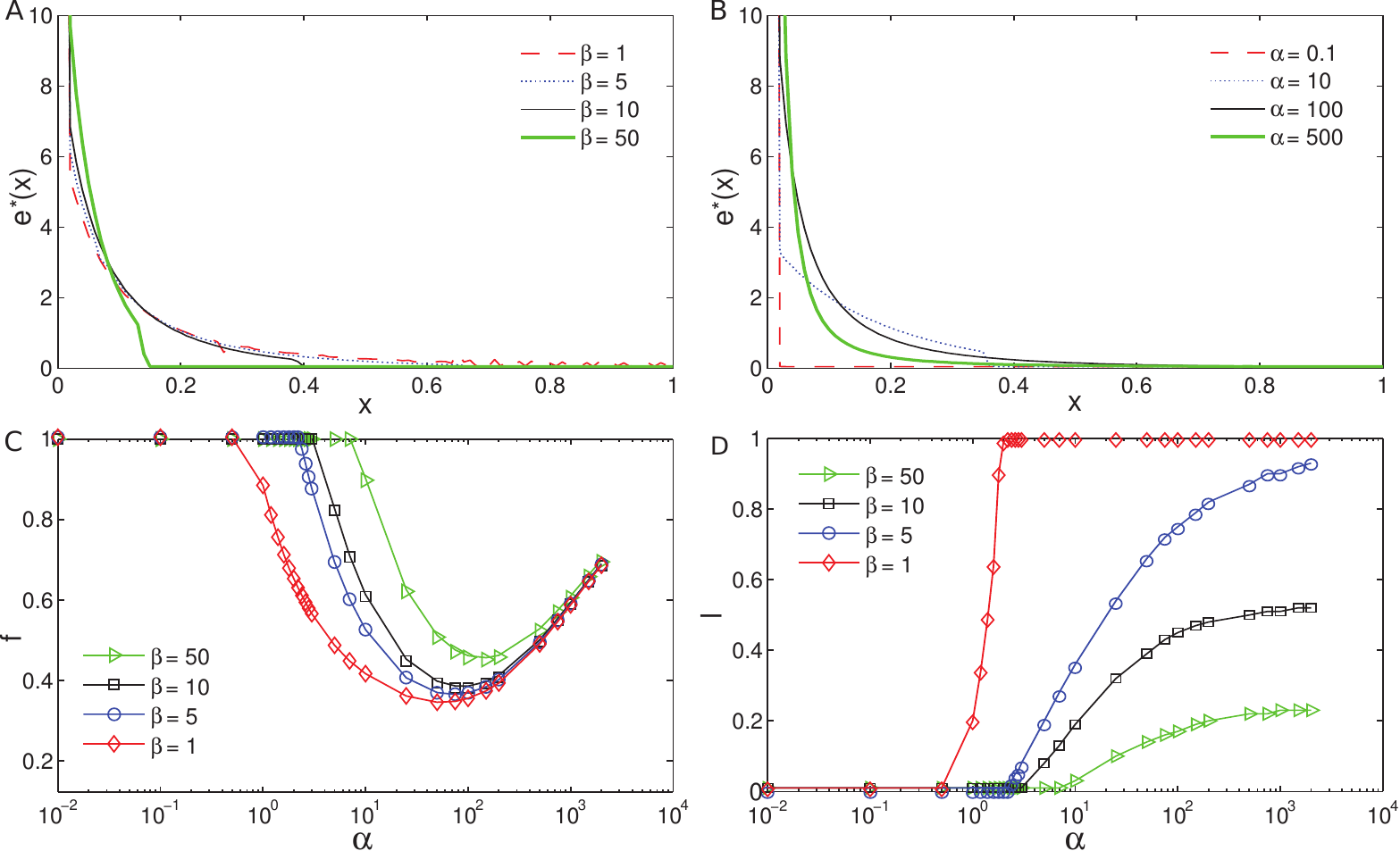}
\caption{(Top) Optimal $E_2$ enzyme profiles as (A) $\beta$ is varied at constant
$\alpha=50$, and (B) $\alpha$ is varied at constant $\beta=5$. All the optimal
profiles are obtained by numerical optimization on a lattice with $N=100$ sites.
(Bottom) Quantifying the optimal profile. (C) The fraction $f$ of $E_2$ enzymes
that are clustered in the optimal profile shows a non-monotonic dependence on
$\alpha$. (D) The extent of distributed enzymes $l$ increases monotonically with
$\alpha$ until reaching a maximal extension that depends on $\beta$.}
\label{fig:leakage_reflecting}
\end{figure*}

The upper panels of Fig.~\ref{fig:leakage_reflecting} show the optimal enzyme
profiles found numerically for different combinations of the parameters $\alpha$
and $\beta$. Importantly, we find that the fully clustered configuration is not
always the optimal distribution; for different parameter values, the optimal 
$E_2$ profile can be either a fully-clustered configuration at $x=0$, or a mixed
profile in which only a finite fraction of the available enzymes are clustered.
This is reminiscent of the behavior of an open system with an absorbing boundary
but without a competing pathway, reported previously \citep{US}, although the
shape of the enzyme profile differs. 

We quantify the level of clustering in the optimal enzyme profiles (see
Fig.~\ref{fig:leakage_reflecting}, lower panels) by the fraction $f$ of $E_2$
that are located at the lattice site $x=0$, and the extent of the optimal
profile via the distance $l$ over which the optimal enzyme density is above a
threshold of $10^{-3}$. Examining first the behavior as $\beta$ is varied
(upper left) we see that for larger $\beta$ the enzyme profile becomes more
concentrated at smaller values of $x$; $f$ increases and $l$ decreases. This is
simply because intermediate molecules typically diffuse less far from the source
before reacting via the secondary pathway. Turning now to the behavior as a
function of $\alpha$, we see that there exists a sharp transition: below a
($\beta$-dependent) critical value of $\alpha$, the optimal profile is the
fully-clustered configuration, $f=1$. As the threshold is crossed, a fraction of
enzymes are relocated away from $x=0$ and distributed over an extended region;
$f$ decreases and $l$ increases. Interestingly, as $\alpha$ is increased further
we find that the available enzymes tend to once again relocate towards $x=0$;
$f$ passes though a minimum and begins to increase again. This is not
accompanied by a decrease in $l$, but the density of enzymes is reduced at
larger $x$ and increased at smaller $x$. For large $\alpha\gtrsim1000$, the
optimal profiles for different values of $\beta$ become more similar (with the
exception of the position at which the enzyme profile cuts off sharply, which
remains $\beta$-dependent).

The pattern of changes in the optimal profile can be understood as follows. When
$\alpha$ is small the reaction efficiency is optimized by a clustered
configuration since this is the enzyme distribution that maximizes the number of
large-$E$ trajectories. However, the clustered configuration also leads to a
large population of trajectories, those corresponding to $I$ molecules that
rapidly diffuse away from the cluster and do not return, with extremely small
values of $E$. At intermediate values of $\alpha$, it becomes favorable to move
some enzymes away from the cluster and to distribute more widely. This increases
the probability of reaction for trajectories that rapidly leave the cluster,
while not significantly reducing the probability of reaction for trajectories
that spend a significant amount of time in the vicinity of the cluster. For
large $\alpha$, however, the need for these distributed enzymes is reduced,
since only for those trajectories that rapidly escape via the secondary pathway
is the reaction probability much less than 1.  Thus, by again concentrating
enzymes around $x=0$ it is possible to maximize the probability of reaction for
those trajectories that spend only a very short time in the system, and
therefore do not diffuse far from $x=0$. Here we see a significant difference
from an open system where loss occurs only at the boundary, for which there is
no impetus to cluster enzymes again \citep{US}. If loss occurs only at $x=1$,
intermediate molecules must always diffuse past all $E_2$ molecules in order to
escape from the system.  However, if loss occurs in the vicinity of $x=0$, then
$E_2$ enzymes placed far from the source are essentially wasted.

\subsubsection{Higher-dimensional systems}

We now consider systems in more than one spatial dimension, beginning with a
three-dimensional spherical geometry. We impose angular symmetry, such that
position within the system can be parametrized by a single radial coordinate,
$r$. We place $E_1$ enzymes at the center $r=0$ of a spherical volume of radius
$1$. In the first instance, we neglect any secondary pathways ($\beta=0$) and
impose an absorbing boundary condition at $r=1$. With these simplifications,
Eq.~\ref{general_dimensionless} becomes
\begin{equation} \label{3d_dimensionless}
	0=r^{-2}\partial_r\left[r^2\partial_r\rho(r)\right]-\alpha e(r)\rho(r)
\end{equation}
with the boundary conditions $[4\pi r^2\partial_r\rho(r)]_{r=0}=-1$, accounting
for the production of intermediate, and $\rho(1)=0$. This system is a
three-dimensional analog of that discussed previously in \citep{US}.

We once again begin by exploring the configurations in which $E_2$ are either
placed on a shell of radius $r_0$, $e_{\rm c}(r)=\delta(r-r_0)/(3r_0^2)$, or
uniformly distributed throughout the spherical volume, $e_{\rm u}(r)=1$ (recall
that these distributions are scaled by the average density such that $V^{-1}\int_V
e(\mathbf{r}){\rm d}\mathbf{r}=1$). The reaction efficiency is then given by 
\begin{equation} \label{3d_cluster_current}
	\left(\frac{J_2}{J_1}\right)_{\rm c}=\frac{\frac{\alpha}{3}(1-r_0)}
		{r_0+\frac{\alpha}{3}(1-r_0)}
\end{equation} 
for the enzyme shell, and 
\begin{equation}
	\left(\frac{J_2}{J_1}\right)_{\rm u}=1-\sqrt{\alpha}\ {\rm csch}\sqrt{\alpha}.
\end{equation}
for uniformly distributed enzymes.

\begin{figure}
\includegraphics{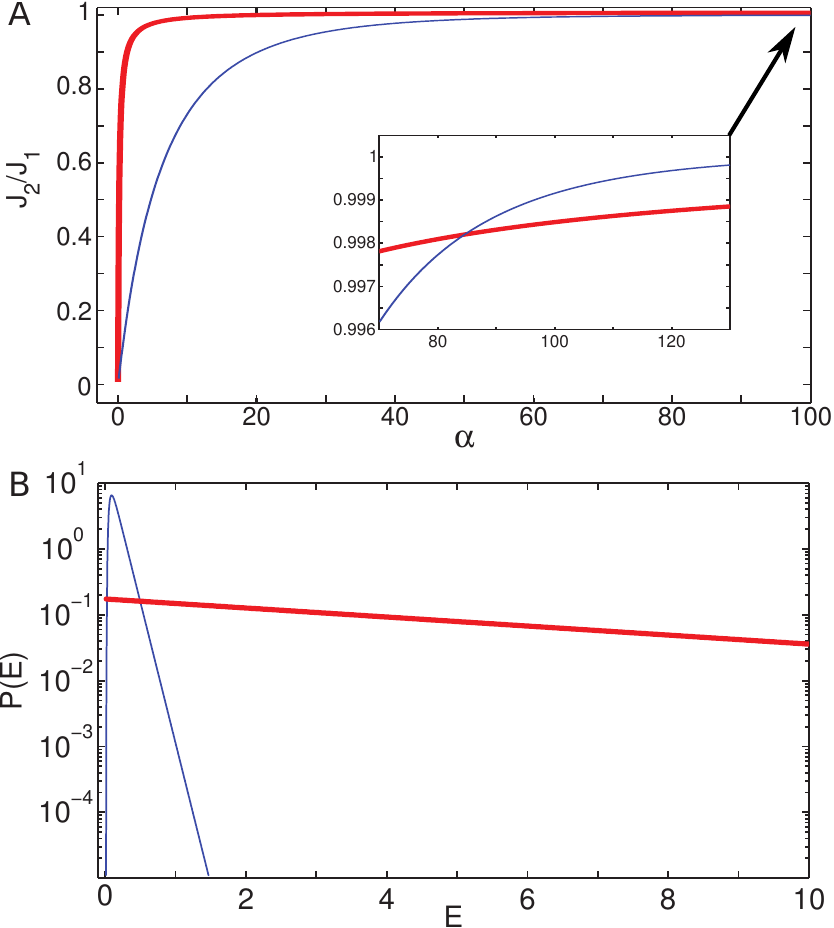}
	\caption{(A) Comparison of the reaction flux for clustered (thick red line, $r_0=0.05$)
and distributed (thin blue line) $E_2$ profiles. The clustered distribution
achieves a significantly higher efficiency for small $\alpha$.  A transition is
observed at $\alpha\approx85$, above which the distributed profile reaches a
marginally higher efficiency.
(B) Enzyme exposure distributions for the clustered (thick red, $r_0=0.05$) and
the distributed (thin blue) $E_2$ profiles. The enzyme exposure distribution for
the distributed $E_2$ profile is sharply peaked at $E\ll1$, whereas the
distribution for the clustered profile extends to large values of $E$.}
	\label{fig:3d_current}
	\label{fig:3d_enzymeexposure}
\end{figure}

As the shell radius $r_0$ approaches zero, the efficiency of reaction of
intermediate with $E_2$ approaches one. However, this situation is not
physically realistic, as enzymes have a finite size and there will be a maximum
packing density which limits the potential shell radii. If $r_0$ is taken to be
small but finite, for small $\alpha$ the clustering of enzymes into a
tightly-packed shell configuration still achieves a higher reaction flux (see
Fig.~\ref{fig:3d_current}A) than the uniformly-distributed configuration.
However above a critical ($r_0$-dependent) $\alpha$ value, the uniform enzyme
arrangement is able to achieve a higher reaction flux.  This transition is
analogous to that seen in the equivalent one-dimensional system \citep{US}.
However, as can be seen in Fig.~\ref{fig:3d_current}A the region in which the
uniform enzyme distribution is favored is shifted to much higher $\alpha$
values, from $\alpha\approx9$ in the one-dimensional system up to
$\alpha\approx85$ for the three-dimensional system with $r_0=0.05$. Since the
transition occurs at extremely large $\alpha$ values, for which almost all
intermediate particles react, the difference between the efficiencies of the two
profiles is extremely small. Furthermore, in the low-$\alpha$ domain the
advantage provided by clustering of enzymes is much more significant than in the
one-dimension system, with an increase in the reaction flux by more than a
factor of four. Thus clustering of enzymes is more strongly favored in
three-dimensional than in one-dimensional systems. 

Calculating the corresponding enzyme exposure distributions, 
\begin{subequations}
	\begin{align}
	P_{\rm u}(E)& =2\sum_{n=1}^{\infty}(-1)^{n+1}(\pi n)^2 e^{-(\pi n)^2E},\\
	P_{\rm c}(E)& =\frac{3r_0}{1-r_0}e^{-\frac{3r_0}{1-r_0}E},
	\end{align}
\end{subequations}
we find that $P_{\rm u}(E)$ is sharply peaked around a small but finite value of
$E$ (see Fig.~\ref{fig:3d_enzymeexposure}B; the mean and variance of $P_{\rm
u}(E)$ are $1/6$ and $1/90$, respectively). For the clustered profile, $P_{\rm
c}(E)$ takes its customary exponential form. Thus we can see that for small
$\alpha\ll1$, almost all of the weight of $P_{\rm u}(E)$ lies in the region
$E\lesssim1$ where $p_r(E)$ is small; the exponential tail of $P_{\rm c}$,
though, results in a larger mean value of $E$ and more trajectories in regions
where $p_r(E)$ is significant.  Thus for $\alpha\ll1$, the clustered
configuration is more efficient. On the other hand when $\alpha\gg1$,
essentially all the probability weight of $P_{\rm u}(E)$ lies in the region
$E\gtrsim\alpha^{-1}$, where $p_r(E)\approx1$; however, $P_{\rm c}(E)$ is
actually largest in the region $E\ll1$ where $p_r(E)$ is small. These
trajectories, which correspond to $I$ molecules that rapidly diffuse away from
the $E_2$ cluster and do not return, generally will not lead to reactions and
thus reduce the relative efficiency of the clustered configuration.

How can the greater impact of clustering in a spherical geometry be understood
intuitively? On a typical path to the boundary, a single intermediate particle
originated form the center explores only a fraction of the whole sphere. If the
same number of enzymes are distributed on a spherical shell further from the
center, the effective ``reaction cross-section'' is smaller because the fraction
of this shell that an intermediate will typically explore decreases, and
with it the fraction of the enzymes in the system to which the intermediate
molecule will be exposed. This is in contrast to the one-dimensional case, where
the particle passes all enzymes before getting absorbed by the boundary. To
derive a benefit from distributing enzymes, a larger $\alpha$ value is therefore
required in three dimensions to compensate for this reduction in the effective 
level of $E_2$ enzymes to which intermediate molecules are exposed.

Next we investigated the optimal enzyme distribution as a function of the
control parameter $\alpha$. As noted above, if the clustering of all $E_2$
enzymes at $r=0$ is permitted then $J_2/J_1\to1$ independent of $\alpha$,
yielding the maximal possible flux. However, we note again that this
configuration with an infinite packing density is not physically realizable.
Instead we impose a limit to the possible packing density through a minimal
radius $r_0$ within which $E_2$ enzymes cannot be placed. We adapt the
optimization procedure described above by solving Eq.~\ref{3d_dimensionless} on
a radial lattice where each lattice site represents a concentric shell of the
system. This fixes a minimal value of $r_{0,{\rm min}}=(2N)^{-1}$ at the
mid-point of the innermost shell; larger values of $r_0$ can be prescribed, but
not smaller shells.

\begin{figure}
	\includegraphics{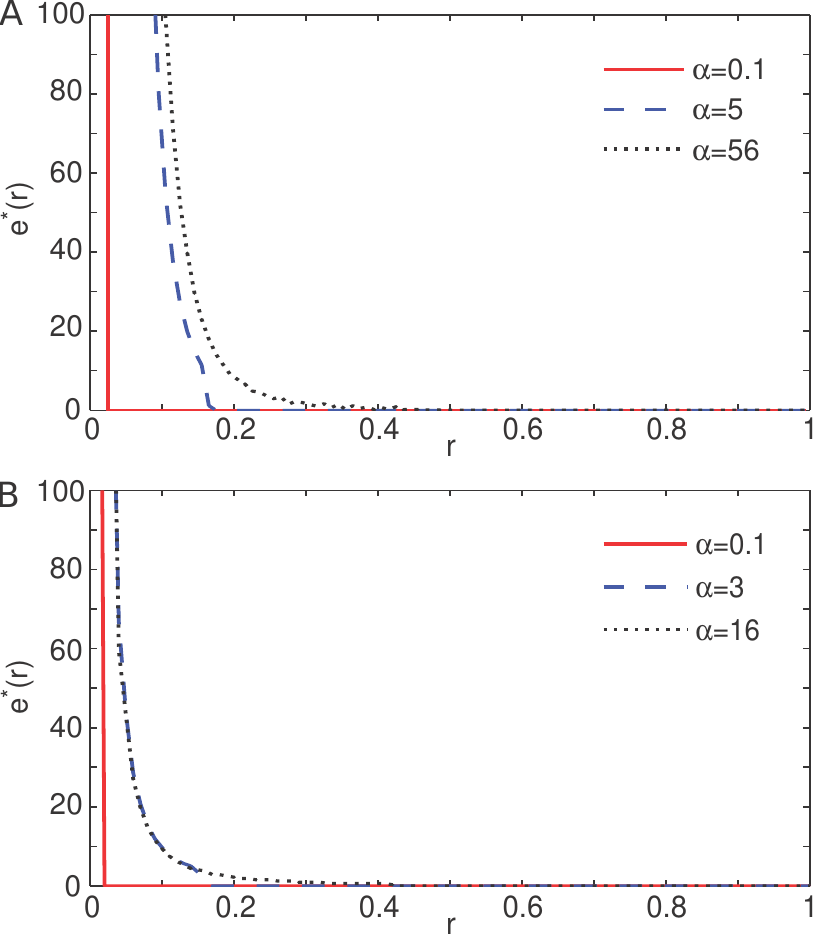}
	\caption{Changes in the optimal enzyme profile as a function of $\alpha$ are
	qualitatively similar in two and three dimensions. (A) For the three-dimensional
	system described by Eq.~\ref{3d_dimensionless}, an extended region of
	distributed enzymes emerges for $\alpha\gtrsim0.05$. 
	(B) In two dimensions, the critical value for the emergence of a distributed
	enzyme fraction is $\alpha\approx0.5$.
	}
	\label{fig:3d_optimal}
	\label{fig:2d}
\end{figure}

As shown in Fig.~\ref{fig:3d_optimal}A, when a finite minimal clustering
radius is imposed we again find that a purely clustered configuration is optimal
for small $\alpha$ while for $\alpha$ larger than a critical value an extended
region of distributed enzymes emerges, in the same way as in one dimension
\cite{US}. The critical $\alpha$ value at which this transition occurs is
strongly dependent on the minimal allowed radius, $r_0$. Interestingly, for
$N=100$ and $r_0=r_{0,{\rm min}}=0.005$ the transition point $\alpha\approx0.05$
is significantly lower than in the one-dimensional system, despite the above
arguments that distributing enzymes is less efficient in three dimensions than
in one; only at $r_0\sim0.2$ does the transition point reach the $\alpha=1$
observed in one dimension. The increased penalty, via the reduction in reaction
cross-section, of moving enzymes to larger values of $r$ is instead reflected in
the shape of the extended enzyme ``tail'': unlike in one dimension, the enzyme
density in the extended region of the profile is not constant but decreases
roughly as $e^*(r)\sim r^{-4}$.

We have seen that one- and three-dimensional systems have qualitatively similar
phenomenology in terms of whether a clustered or uniform profile is preferable
and in terms of the optimal enzyme profile, although the quantitative aspects of
these transitions vary. The same also holds true for the two-dimensional case:
we find that this system also displays a transition from a clustered to uniform
$E_2$ configuration being preferable at an $\alpha$ value between those at which
the transition occurs in one- and three-dimensional systems (data not shown).
Figure~\ref{fig:2d}B shows that once again an extended optimal profile appears
at $\alpha\approx0.5$ (for $r_0=0.005$; the effect of varying $r_0$ is much
weaker in two dimensions than in three). In the extended region, the optimal
density is found to decrease as $e^*(r)\sim r^{-2}$. We can therefore see that
the underlying physics of these transitions is generic, and is not dependent on
the statistics of diffusion in particular dimensions.

\subsection{Non-linear reactions} \label{sec:nonlinear}

We now turn to the case of fully non-linear reactions. We begin by considering
the case $\beta=0$, in which there is no competition with secondary pathways
for the intermediate $I$. We furthermore restrict ourselves to a one-dimensional
system on the domain $x\in[0,1]$, with $E_1$ enzymes located at $x=0$ and an
absorbing boundary condition at $x=1$. That is, we consider a reaction-diffusion
equation of the form
\begin{equation} \label{nonlin_dimensionless}
	0=\partial_x^2\rho(x)-\frac{\alpha e(x)\rho(x)}{1+\gamma\rho(x)},
\end{equation}
together with source-sink boundary conditions $\partial_x\rho\vert_{x=0}=-1$ and
$\rho(1)=0$, where $\gamma$ is the effective saturation parameter defined in
the `Model' Section. 

\begin{figure*}
\includegraphics{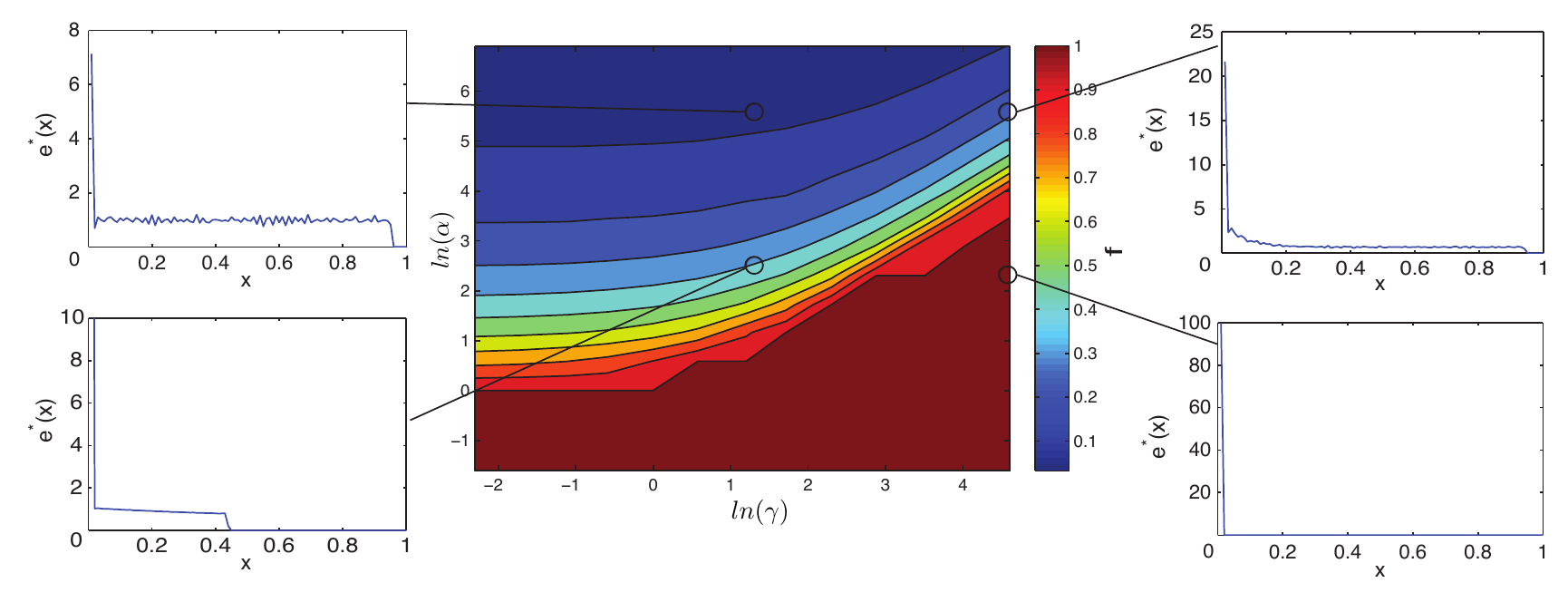}
	\caption{Central panel: the fraction $f$ of $E_2$ enzymes that are clustered in the
numerically-optimized enzyme distribution (on a lattice of 100 sites) as the
parameters $\alpha$ and $\gamma$ are varied. Outer panels depict the optimal
profiles at specific parameter combinations, showing how the profile shape
changes across the transition from clustered to fully-distributed.
Optimizations were run for $2\times10^5$ steps on a lattice of $N=100$ sites,
with a solution error-tolerance of $10^{-7}$.
}
\label{fig:nonlinear}
\end{figure*}

We again seek to find the $E_2$ distribution $e(x)$ that maximizes the reaction
efficiency $J_2/J_1$ via numerical optimization. The non-linear nature of the
reaction terms mean that the discretized reaction-diffusion equation for a given
$e(x)$ no longer takes the form of a linear system that can be solved directly.
Instead, we used a shooting approach to calculate $\rho(x)$ and thereby $J_2$.
An initial trial solution for $\tilde{\rho}(x_N)$ at the right-most lattice site
is selected. This trial is then used to successively solve the non-linear
reaction-diffusion equation at the remaining lattice sites (the equation for
site $N$ depends on $\rho(x_N)$ and $\rho(x_{N-1})$, that for site $N-1$ depends
on $\rho(x_N)$, $\rho(x_{N-1})$ and $\rho(x_{N-2})$, and so on). Once a trial
solution $\tilde\rho(x)$ has been calculated for all sites, this is tested
against the reaction-diffusion equation at site $x_1$, which includes the source
boundary condition. If the equation is satisfied to within a certain tolerance,
then the solution is accepted. Otherwise, the trial value of $\tilde\rho(x_N)$
is refined, and the process repeated. The mutation, selection and mixing steps
of the optimization were unchanged. 

Figure~\ref{fig:nonlinear} shows results for the optimal enzyme profiles 
for different values of $\alpha$ and $\gamma$. We first
verified that this solution technique accurately reproduces the results of the
linear system in the limit of small $\gamma$, which should correspond the
results for the linear-reaction case that have been described previously
\citep{US}. Indeed, we find that when $\alpha<1$ all $E_2$ enzymes should be
co-localized with the $E_1$ enzymes at $x=0$. As $\alpha>1$ is increased, the
fraction $f$ of $E_2$ enzymes that cluster at $x=0$ decreases with the remaining
enzymes being distributed uniformly over an extended region such that $e(x)=1$
in this region. Figure~\ref{fig:nonlinear} shows that the same
qualitative behavior is also observed for larger values of $\gamma$.  For all
values of $\gamma$ tested, the optimal profile undergoes a transition from fully
clustered at small $\alpha$ to a mixed profile with a clustered fraction and
extended, lower-density, region for larger $\alpha$. The critical $\alpha$ value
at which this transition occurs increases with $\gamma$, since in the fully
clustered configuration a larger $\gamma$ serves to reduce the effective
reaction rate. Additionally, we find that the optimal profiles deviate in shape
and extension away from the source. In the mixed-profile regime, the extended
``tail'' of enzymes need not have a constant density. This reflects the fact
that for intermediate values of $\alpha$ the level of saturation of $E_2$
enzymes will vary with position. Finally, it appears that once the transition to
a mixed profile has begun, the fraction of $E_2$ enzymes clustered at $x=0$
decreases more quickly as $\alpha$ is increased if $\gamma$ is large than if
$\gamma$ is small. We attribute this to the fact that large values of $\alpha$
tend to dramatically reduce the intermediate density within the system, thereby
moving a system that was in the saturated regime for small $\alpha$ into the
unsaturated regime for large $\alpha$. Indeed, for extremely large values of
$\alpha$ the optimal profile becomes independent of $\gamma$ and approaches that
expected in the linear reaction case.

The inclusion of non-linear reaction terms complicates the analysis of systems
of this type via enzyme exposure. This is because the probability of reaction of
an individual $I$ molecule depends not only on the enzyme density, but also on
other intermediate molecules in the system. One can define the effective enzyme
activity at each position as $e(x)/\left[1+\gamma\rho(x)\right]$, and thereby
calculate an effective enzyme exposure for a trajectory as $\tilde
E=\int_0^{t_{\rm escape}} e(x(t))/\left[1+\gamma\rho(x(t))\right]{\rm d}t$
taking $\rho(x)$ to be the solution to Eq.~\ref{nonlin_dimensionless}. However,
it must be noted that this does not lead to a true decomposition of the reaction
flux into diffusion- and reaction-dependent terms because $\rho(x)$ itself
depends on the reaction parameter $\alpha$.

\begin{figure}
\includegraphics{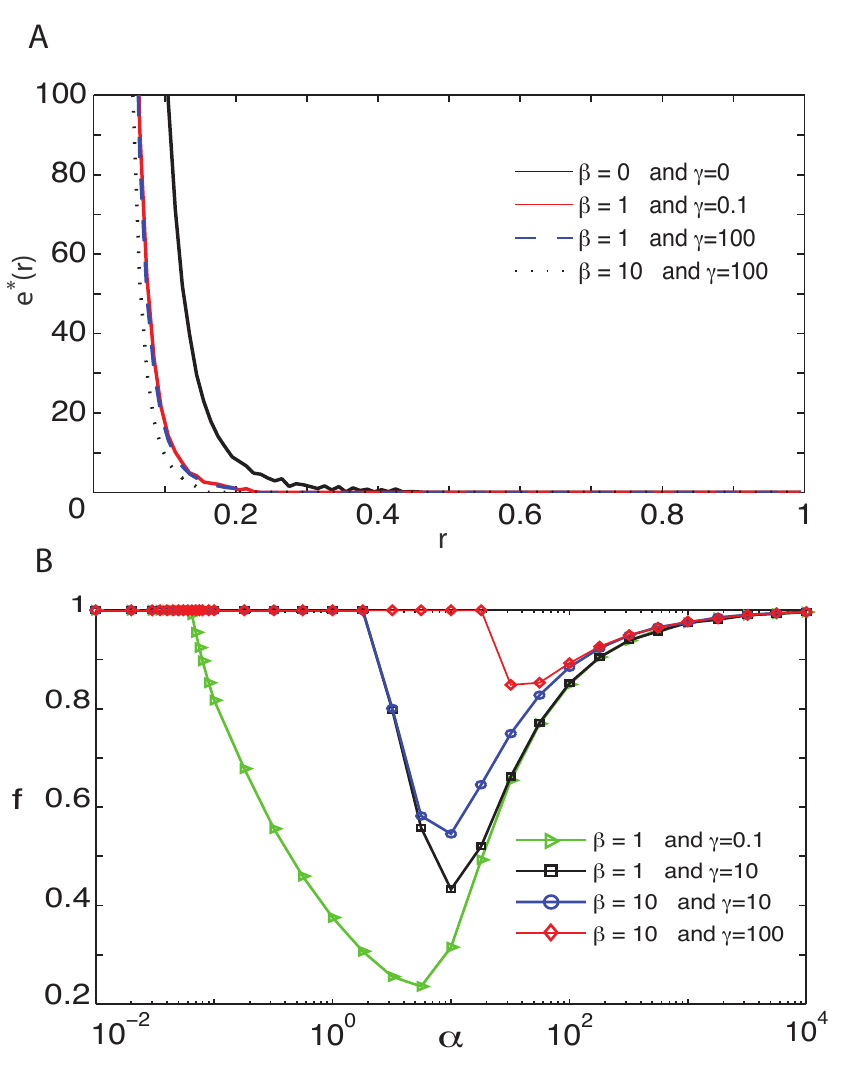}
\caption{
	(A) Optimal enzyme distributions for the full model, with parameter values
		as indicated, with $\alpha=56$.
	(B) Clustered fraction of $E_2$ enzymes in the optimal profile $f$, as a
		function of the catalytic activity of $E_2$. A non-monotonic dependence is
		seen when $\beta\neq0$.}
\label{fig:full}
\end{figure}

Finally, we briefly consider the results of the full model described in Section
\ref{sec:Model}, including both a non-linear reaction, $\gamma\neq0$, and a 
competing pathway, $\beta\neq0$, in a radially-symmetric
three-dimensional geometry.  Figure~\ref{fig:full}A shows that the optimal
enzyme distributions are qualitatively similar to those in
Fig.~\ref{fig:3d_optimal}A for a system without competition and with only linear
reactions. However, examining the fraction of enzymes that are clustered at
$r_0$ (Fig.~\ref{fig:full}B), we see that $f$ is always larger than in the
limiting case $\beta,\gamma\to0$. This is consistent with our results above for
one-dimensional systems with only competition or only non-linearity in the
reaction with $E_2$ enzymes, where we found that increasing the strength of
either of these effects will increase the tendency for clustering of $E_2$
enzymes. The non-monotonic dependence of $f$ on $\alpha$ that was previously
found (Fig.~\ref{fig:leakage_reflecting}C) when only competition is present in
the model is also preserved in the case $\gamma\neq0$. In a similar way to
Fig.~\ref{fig:nonlinear}, Fig.~\ref{fig:full}B also suggests that the principle
effect of varying $\gamma$ is to alter the threshold value of $\alpha$ beyond
which the purely clustered profile becomes sub-optimal. In summary, these
results indicate that the qualitative effects of each of the modifications that
we have previously considered individually are representative of the impact of
the same elements in the full model.

\section{Discussion}

In this work we have demonstrated that the existence of a transition between
clustered and distributed optimal arrangements of enzymes is a generic feature
of diverse reaction-diffusion systems. While the exact shape of the optimal
enzyme profile and the parameter-dependence of this transition varies with the
specific system, the fact that a non-trivial optimal profile exists is a general
result of the interplay of reaction and diffusion in such systems. The
transition ultimately emerges from the stochastic dynamics of individual
intermediate molecules, as demonstrated by its dependence on the interplay
between the distributions of enzyme exposure and reaction probability. By
examining these distributions, we are led to an intuitive explanation for the
transition. When reactions are slow, clustering of enzymes is beneficial because
this provides the highest enzyme density in the region in which intermediate
molecules are most likely to spend a significant amount of time. When reactions
become fast, a limited density of enzymes will already ensure the rapid reaction
of these molecules; in this scenario, it becomes preferable to distribute a
fraction of enzymes more widely so as to provide an opportunity to react with
those intermediate molecules that rapidly escape from the enzyme cluster.

For specific systems, our analysis has revealed some further notable features.
In systems with competing pathways, the optimal enzyme profile tends to
concentrate near the source again as the reaction rate is increased further,
which results from the fact that intermediates can be lost from the region near
the source as opposed to only at the boundary of the system. We have also seen
that the benefit of clustering increases with the effective dimension of the
system, as the increased space available to diffusive trajectories means that
intermediates typically only have access to a small fraction of distributed 
enzymes. Finally, our results for non-linear reactions suggest that clustering
again enhances pathway efficiency if the availability of intermediates 
(determined by the activity of the first enzyme in the pathway) is increased. 
This observation suggests that it may be desirable to dynamically regulate the 
localization of enzymes, and specifically the formation of multi-enzyme
complexes, in response to the availability of substrate or the flux of upstream
reactions. A potential example of such regulation is provided by mammalian
hexokinase isoform HKII, which is thought to undergo reversible translocation
between the outer membrane of mitochondria and a more diffuse cytoplasmic
distribution depending on factors including glucose-6-P and GSK3
\cite{Wilson78, Ribalet11}, thereby altering the relative flux of glucose through
different metabolic pathways.

While there are several well-documented examples of enzyme clustering, including
the pyruvate dehydrogenase and cellulosome complexes mentioned above
\cite{deKok1998, Bayer1998} as well as glycolytic enzymes in various cell types
\cite{Sullivan03, Anderson05, Campanella05}, enzyme clustering is not thought to
be the default strategy in molecular biology. For example, while the cellulosome
is a conglomeration of enzymes tethered to the outside of bacteria, other enzyme
classes such as proteases \cite{Wandersman89} do not typically form tethered
complexes but rather are simply secreted into the extracellular environment. We
are not aware of specific enzyme systems that display a  combination of a
cluster with a more diffuse arrangement near the cluster.  Observation of such
localization patterns will be difficult, due to the relatively low density and
dynamic nature of the distributed region in close proximity to the high-density
cluster. Such enzyme distributions could be generated with the help of
pre-existing cellular structures such as the cytoskeleton \cite{Masters84} or
membrane sub-domains \cite{Simons04}. Simpler arrangements consisting of a
localized and a uniform fraction would naturally arise from weak, transient
interactions between the enzymes.  

Our principal conclusions could be tested experimentally using, for example, the 
``single-molecule cut-and-paste'' technique \citep{Kufer08} or
DNA origami \cite{Rothemund06, Douglas2009, Wilner2009} to construct specific
arrangements of enzymes. Such constructs could also be incorporated into 
microfluidic chambers featuring localized sources and sinks of substrate.
The reaction efficiency could be measured by observing the relative quantities
of reacted and un-reacted substrate in the efflux channel, or using 
fluorogenic substrates.  Such experiments would allow for quantification of
the relative reaction efficiencies of different enzyme arrangements as the
parameter $\alpha$ is varied by altering, for example, the number of enzymes in
the system or the substrate diffusion constant. 

While the continuous reaction-diffusion models that we have considered here
present a useful mesoscopic description of the enzymatic systems, they make a
number of approximations that will limit their validity at extremely short
length scales.  Foremost amongst these is that neither enzymes nor intermediate
molecules occupy any finite volume. In reality there will be an upper limit to
the number of enzymes that can be clustered within a certain region.
Furthermore, steric hindrance by enzymes will affect the trajectories of
intermediate molecules. Thus the tight packing of enzymes may strengthen the
effect of clustering by physically blocking the escape of intermediates.
However, by the same measure, a tight clustering of enzymes may prevent the
access of initial substrates into the cluster. At such short length-scales, it
is also not clear to what extent the motion of intermediates can be represented
as normal diffusion. Additionally, enzymes are not reactive over their entire
surfaces but only at specific catalytic binding sites. While rotational
orientation can generally be neglected for freely-diffusing enzymes, these
effects may become significant if enzymes are attached to rigid scaffolds. A
more complete understanding of these issues, together with more complex reaction
schemes including cooperativity and allosteric regulation, will be crucial for a
complete understanding of the design principles underlying enzyme arrangements
in living cells as well as the effective engineering of synthetic biochemical
systems.

\acknowledgments
This research was supported by the German Excellence Initiative via the program
`Nanosystems Initiative Munich' and the German Research Foundation via the SFB
1032 `Nanoagents for Spatiotemporal Control of Molecular and Cellular
Reactions' and via the Focus area SPP 1617. FT is supported by a research
fellowship from the Alexander von Humboldt Foundation.

\section*{Appendix: Competing pathway with absorbing boundary conditions}
\label{app_leakage}

Here we briefly consider a system described by 
Eq.~\ref{reaction_diffuison_leakage}, but with an absorbing boundary condition
$\rho(1)=0$ rather than the reflecting boundary considered in section
\ref{leakage}.  Comparing the known results for the limit $\beta\to0$
\cite{US}, and for the reflecting boundary condition discussed above, we can
predict that there should be a qualitative difference in whether clustered or
uniform profiles are preferred as $\beta$ is varied. For small $\beta\ll1$, loss
to the competing pathway will be negligible compared to loss through the
boundary at $x=1$. We would therefore expect that as $\alpha$ is increased, the
system should undergo a transition from a regime in which the clustered
configuration is preferable to a regime in which the uniform profile provides a
higher efficiency. On the other hand, for large $\beta$ the length scale
associated with the loss to the secondary pathway is short compared to the
system size. In this case, the choice of boundary condition of $x=1$ should have
little influence on the dynamics, which should resemble that described in
section \ref{leakage} where the clustered configuration is always preferable.

\begin{figure}
\includegraphics{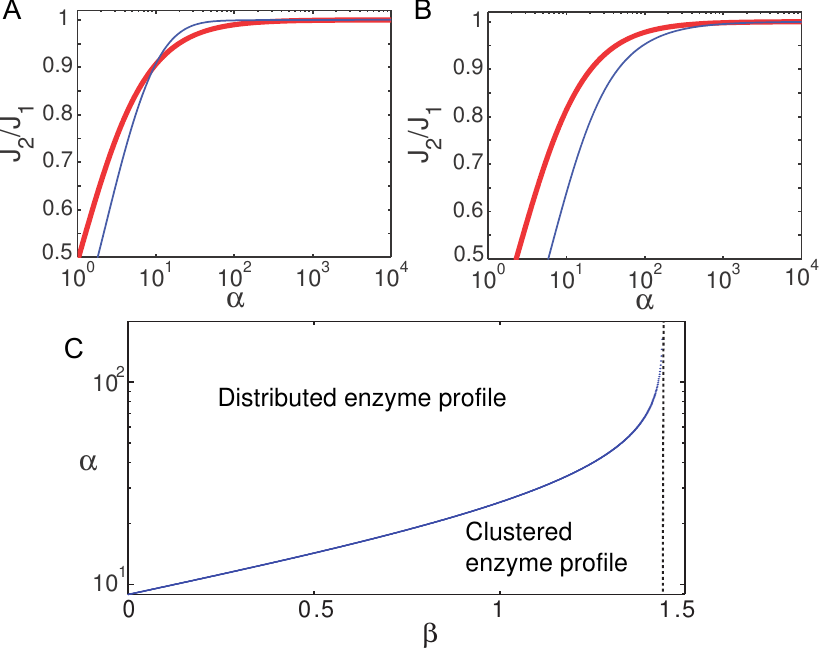}
\caption{(A,B) Comparison of the reaction efficiency for clustered (thick red line) and
distributed (thin blue line) $E_2$ enzymes reveals a qualitatively different
behavior in different $\beta$ ranges. For $\beta=0.1$ (A) a cross-over is
observed at which the profile with the higher efficiency changes. For $\beta=5$
(B) clustering is preferable for all values of $\alpha$.
(C) The transition value of $\alpha$ dividing the regimes where the clustered or
uniform profiles achieve a higher reaction flux, found by solving
$(J_2/J_1)_{\rm u}=(J_2/J_1)_{\rm c}$ numerically. Above the critical $\beta_c
\approx 1.4$ (dashed line) the transition disappears and the clustering of
enzymes is always more efficient.}
\label{fig:leakage_absorbing}
\end{figure}

The reaction fluxes obtained by solving this system for uniform and clustered
enzyme configurations are 
\begin{align} \label{current_delta_absorbing_1d}
	\left(\frac{J_2}{J_1}\right)_{\rm u}& =\frac{\alpha}{\alpha+\beta}\left(1-{\rm
		sech}\sqrt{\alpha+\beta}\right), \\
	\label{current_uniform_absorbing_1d}
	\left(\frac{J_2}{J_1}\right)_{\rm c}& =\frac{\alpha}
	 {\alpha+\sqrt{\beta}\ {\rm coth}\sqrt{\beta}}.
\end{align}
Figure~\ref{fig:leakage_absorbing}A and B confirms that there is a difference in
whether the clustered or uniform profile is more efficient in different regimes
of $\beta$, in keeping with our expectations.
Figure~\ref{fig:leakage_absorbing}C plots the critical value of $\alpha$ as a
function of $\beta$, and demonstrates that the transition disappears at a finite
value of $\beta_c\approx1.4$. When $\beta>\beta_c$ the clustered configuration
always achieves a higher reaction flux.

\newpage
\widetext
\setcounter{equation}{0}
\renewcommand{\theequation}{S\arabic{equation}}
\section*{SUPPLEMENTARY MATERIAL}
\setcounter{section}{0}
\renewcommand{\thesection}{S\arabic{section}}
\section{Equivalence of expressions for the reaction efficiency}

The reaction-diffusion equation leads us to define the reaction efficiency 
according to the Eq.~\ref{conservation_equation} of the main text, 
\begin{equation} \label{eq:efficiency1}
	\frac{J_2}{J_1}=\int \alpha
	e(\mathbf{r})\frac{\rho(\mathbf{r})}{1+\gamma\rho(\mathbf{r})} {\rm d}\mathbf{r}.
\end{equation}
Here we show that the alternative expression (Eq.~\ref{eq:efficiency_traj} of the main text) 
\begin{equation} \label{eq:efficiency2}
	\frac{J_2}{J_1}=\int_0^\infty P(E)p_{\rm r}(E){\rm d}E,
\end{equation}
which arises from the examination of individual intermediate trajectories, can
be derived from Eq.~\ref{eq:efficiency1} in the linear regime where $\gamma\to0$.

We begin by reformulating the steady-state density $\rho({\mathbf r})$ in terms
of trajectories of diffusing molecules of intermediate. We denote the diffusive
trajectory of a single intermediate molecule, in the absence of any $E_2$
enzymes, as $\rt$. Such a trajectory has associated with it a time $t_{\rm
escape}$ after which the trajectory is terminated, either by escape across the
system boundary or loss to a secondary pathway. The reintroduction of $E_2$
enzymes according to the distribution $e({\mathbf r})$ leads to an instantaneous
propensity for conversion to the correct product at each point along the
trajectory of $\alpha e(\mathbf{r}(t))$. Thus the survival probability
$S(t|\rt)$ that an intermediate molecule on the trajectory $\rt$ has not
undergone a reaction with $E_2$ before the time $t$ follows
$\dot{S}(t|\rt)=-\alpha e(\mathbf{r}(t))S(t|\rt)$. This equation can be
integrated to yield 
\begin{equation} \label{eq:survival}
	S(t|\rt)=\exp\left[-\alpha\int_0^t{\rm d}t'\ e(\mathbf{r}(t'))\right], \ \
	t\leq t_{\rm escape}.
\end{equation}

At steady state, each trajectory $\rt$ of an intermediate molecule will make a
contribution to the total intermediate density at point $\mathbf{r}$ that
depends on the total time that the trajectory spends at $\mathbf{r}$, weighted
by the probability that the intermediate molecule has not yet undergone a
reaction prior to each return to $\mathbf{r}$. This later weighting factor is
simply the survival probability $S(t|\rt)$. Therefore, the local enzyme density
can be rewritten as 
\begin{equation} \label{eq:redensity}
	\rho(\mathbf{r})=\int {\rm d}\rt\ p(\rt)
	 \int_0^{t_{\rm escape}}{\rm d}t\ S(t|\rt)\delta\left[\mathbf{r}-\mathbf{r}(t)\right], 
\end{equation}
where the inner integral is the weighted time spent by a single trajectory at
$\mathbf{r}$, and the outer integral sums over the contributions of all possible
trajectories weighted by the probability $p(\rt)$ of a specific trajectory
$\rt$ occurring. Substituting Eqs.~\ref{eq:redensity} and \ref{eq:survival} into
Eq.~\ref{eq:efficiency1} and changing the order of integration, we find
\begin{align}
	\frac{J_2}{J_1}& =\int{\rm d}\rt\ p(\rt)\int_0^{t_{\rm escape}}{\rm d}t
	\int{\rm d}\mathbf{r}\ \alpha e(\mathbf{r}) e^{-\alpha\int_0^t{\rm d}t'\
		e(\mathbf{r}(t'))}
	\delta\left[\mathbf{r}-\mathbf{r}(t)\right]  \nonumber \\
	& =\int{\rm d}\rt\ p(\rt)\int_0^{t_{\rm escape}}{\rm d}t\ 
		\alpha e(\mathbf{r}(t)) e^{-\alpha\int_0^t{\rm d}t'\
		e(\mathbf{r}(t'))} \nonumber \\
	& =\int{\rm d}\rt\ p(\rt)\int_0^{t_{\rm escape}}{\rm d}t\ 
	\frac{\rm d}{{\rm d}t}\left[-e^{-\alpha\int_0^t{\rm d}t'\ 
		e(\mathbf{r}(t'))}\right] \nonumber \\
	& =\int{\rm d}\rt\ p(\rt)\left[1-e^{-\alpha\int_0^{t_{\rm escape}}{\rm d}t'\ 
		e(\mathbf{r}(t'))}\right].
\end{align}
Finally, defining $E=\int_0^{t_{\rm escape}}{\rm d}t\ e(\mathbf{r}(t))$ we can change the
variable of integration from $\rt$ to $E$, recovering
\begin{equation}
	\frac{J_2}{J_1}=\int_0^\infty P(E)(1-e^{-\alpha E}){\rm d}E.
\end{equation}

\section{One dimension including a competing pathway}

We consider the rescaled reaction diffusion equation as stated in the main text
\begin{equation}
0=\partial x^2 \rho(x)-\alpha e(x)\rho(x)-\beta \rho(x),
\label{appendix_1d_competing_pathway}
\end{equation}
with the source boundary conditions $\partial_x \rho(x)|_{x=0}=-1$. Below we
consider the cases of reflecting ($\partial_x \rho(x)|_{x=1}=0$) and absorbing
($ \rho(x)|_{x=1}=0$) boundaries at $x=1$. 

\subsection{Reflecting boundary, $\partial_x \rho(x)|_{x=1}=0$}
\subsubsection{Clustered enzyme profile}\label{clu_enzyeme_section}

The enzyme profile is taken to be clustered at some point $x_0$,
$e_{\rm c}(x)=\delta(x-x_0)$. In the end we take the limit $x_0$ goes to zero,
leading to a clustering of $E_2$ at the origin. We divide the system into two
parts, part $I$ where $x < x_0$ and part $II$ where $x > x_0$. In each part
Eq.~\ref{appendix_1d_competing_pathway} reduces to 
\begin{equation}
0=\partial x^2 \rho(x)-\beta \rho(x),
\end{equation} 
which has the solution
\begin{equation}
\rho_{i}(x)=A_{i}e^{\sqrt{\beta}x}+B_{i}e^{-\sqrt{\beta}x}
\end{equation}
with $i=\{I,II\}$. Applying the boundary conditions at $x=0$ and $x=1$ yields
\begin{equation}
A_I-B_I=(\sqrt{\beta})^{-1} \qquad A_{II}e^{\sqrt{\beta}}-B_{II}e^{-\sqrt{\beta}}=0.
\end{equation} 
In order to determine all constants we impose two additional conditions, firstly
the matching condition of the concentration of intermediates at $x_0$,
$\rho_{I}(x_0)=\rho_{II}(x_0)$ leading to
\begin{equation}
A_{I}e^{\sqrt{\beta}x_0}+B_{I}e^{-\sqrt{\beta}x_0}=A_{II}e^{\sqrt{\beta}x_0}+B_{II}e^{-\sqrt{\beta}x_0}
\end{equation}
The second condition, which captures particle conservation in the system, is
found by integrating Eq.~\ref{appendix_1d_competing_pathway} from
$x_0-\epsilon$ to $x_0+\epsilon$ and taking the limit of small $\epsilon$,
\begin{equation}
\lim_{\epsilon \to
0}\bigg((\partial_x\rho_{II}(x))_{x_0+\epsilon}-(\partial_x\rho_I(x))_{x_0-\epsilon}-\alpha\rho(x_0)-\beta
\int_{x_0-\epsilon}^{x_0+\epsilon} \rho(x) dx \bigg)=0.
\end{equation} 
The last term on the left hand side vanishes in the limit $\epsilon \rightarrow
0$. This leads to the expression
\begin{equation}
\bigg [A_{II}e^{\sqrt{\beta}x_0}-B_{II}e^{-\sqrt{\beta}x_0} - \bigg(A_{I}e^{\sqrt{\beta}x_0}+B_{I}e^{-\sqrt{\beta}x_0} \bigg)\bigg]
-\frac{\alpha}{\sqrt{\beta}}\bigg(A_{I}e^{\sqrt{\beta}x_0}+B_{I}e^{-\sqrt{\beta}x_0} \bigg)=0. 
\end{equation}
Calculate the reaction current yields
\begin{equation}
\frac{J_2}{J_1}=\alpha \int_{0}^{1}\delta(x-x_0)\rho(x)dx=\alpha(A_{II}e^{\sqrt{\beta}x_0}+B_{II}e^{-\sqrt{\beta}x_0})
\end{equation} 
After some straightforward algebra we arrive at expressions for all four
constants $\{A_{I},A_{II},B_{I},B_{II}\}$. Last we plug them into the above
equation and take the limit $x_0 \rightarrow 0$,
\begin{equation}
	\bigg(\frac{J_2}{J_1}\bigg)_{\rm c}=\frac{\alpha}{\alpha+\beta^{\frac{1}{2}}\tanh(\beta^{\frac{1}{2}})}.
\label{appendix_1d_ref_cluster}
\end{equation}

\subsubsection{Uniform enzyme profile}\label{uni_enzyme_section}

The reaction diffusion equation with a uniform enzyme profile $e_{\rm u}(x)=1$
reads
\begin{equation}
0=\partial x^2 \rho(x)-(\alpha +\beta )\rho(x).
\end{equation} 
The solution is given by
\begin{equation}
\rho(x)=Ae^{\sqrt{\alpha+\beta}x}+Be^{-\sqrt{\alpha+\beta}x}
\end{equation}
Applying the boundary conditions at $x=0$ and $x=L$ leads to the conditions
\begin{equation}
A-B=(\sqrt{\alpha+\beta})^{-1} \qquad Ae^{\sqrt{\alpha+\beta}}-Be^{-\sqrt{\alpha+\beta}}=0.
\end{equation} 
Similarly to above, the constants $A$ and $B$ can be obtained straightforwardly,
and the efficiency of the pathway is
\begin{equation}
	\bigg(\frac{J_2}{J_1}\bigg)_{\rm u}= \alpha \int_0^1 \rho(x)dx=\frac{\alpha}{\alpha+\beta}
\label{appendix_1d_ref_uni}
\end{equation}

\subsection{Absorbing boundary, $\rho(x)|_{x=1}=0$}

The approach is very similar to the one the preceding section. The only
difference, however, is that the boundary condition $ \rho(x)|_{x=1}=0$ leading
to a slightly different second condition for the clustered enzymes
\begin{equation}
A_{II}e^{\sqrt{\beta}}+B_{II}e^{-\sqrt{\beta}}=0.
\end{equation} 
Likewise, we obtain 
\begin{equation}
Ae^{\sqrt{\alpha+\beta}}-Be^{-\sqrt{\alpha+\beta}}=0
\end{equation}
as the second boundary condition for the uniformly distributed enzymes.
Similarly, to the approach in section \ref{clu_enzyeme_section} and  \ref{uni_enzyme_section}, respectively
the corresponding efficiency for the clustered case is given by
\begin{equation}
\bigg(\frac{J_2}{J_1}\bigg)_c=\frac{\alpha}{\alpha+\beta^{\frac{1}{2}}\coth(\beta^{\frac{1}{2}})}
\label{appendix_1d_abs_cluster}
\end{equation}
and for the uniform case
\begin{equation}
\bigg(\frac{J_2}{J_1}\bigg)_u=\frac{\alpha}{\alpha+\beta}(1-\text{sech}(\sqrt{\alpha+\beta})).
\label{appendix_1d_abs_uni}
\end{equation}
 
\section{Enzyme exposure probability distribution in 1D}\label{enzyme_exp_1d_section}

As shown in the main text, the efficiency of the pathway in terms of the enzyme
exposure probability distribution $P(E)$ reads
\begin{equation}
\frac{J_2}{J_1}=1-\int_0^{\infty}P(E)e^{-\alpha E}dE=1-\frac{J_{\rm{loss}}}{J_1}.
\label{appendix_1d_enzyme_exposure}
\end{equation}
To obtain an exact expression of $P(E)$ for the respective case it is convenient
to calculate the inverse Laplace transformation of $J_{\rm{loss}}/J_1$ with
respect to $\alpha$. We have seen above that the expression for the efficiency
often has the form $\alpha/(\alpha+f(\beta))$; hence the loss term is
$f(\beta)/(\alpha+f(\beta))$. The inverse Laplace transformation is easily
obtained and has the form
\begin{equation}
P(E)=f(\beta)e^{-f(\beta)E}.
\end{equation}
For the case of a reflecting outer boundary we therefore have for the uniform
enzyme profile, $f(\beta)=\beta$ and thus $P_{\rm u}(E)=\beta e^{-E\beta}$; for
the clustered enzyme profile,
$f(\beta)=\beta^{\frac{1}{2}}\tanh\beta^{\frac{1}{2}}$ and $P_{\rm
c}(E)=\beta^{\frac{1}{2}}\tanh\beta^{\frac{1}{2}}
e^{-E\beta^{\frac{1}{2}}\tanh\beta^{\frac{1}{2}}}$.

In the case of an absorbing boundary, the clustered distribution also gives rise
to a similar expression for the efficiency, with
$f(\beta)=\beta^{\frac{1}{2}}\coth\beta^{\frac{1}{2}}$ and thus $P_{\rm
c}(E)=\beta^{\frac{1}{2}}\coth\beta^{\frac{1}{2}}e^{-E\beta^{\frac{1}{2}}\coth\beta^{\frac{1}{2}}}$.
However, with a uniform distribution the loss flux does not have the form
discussed above, and thus the calculation of the inverse Laplace transformation
is more involved. We rewrite Eq.~\ref{appendix_1d_abs_uni} and use
Eq.~\ref{appendix_1d_enzyme_exposure} to obtain
\begin{equation}
\bigg(1-\frac{J_2}{J_1}\bigg)=\frac{J_{\rm{loss}}}{J_1}=\frac{\alpha}{(\alpha+\beta)\cosh(\sqrt{\alpha+\beta})}
+\frac{\beta}{\beta+\alpha}.
\end{equation} 
Where the second term has the form as we have already discussed above, thus
\begin{equation}
	P_{\rm u}(E)={\mathcal
L}^{-1}\left[\frac{\alpha}{(\alpha+\beta)\cosh(\sqrt{\alpha+\beta})}\right]+\beta
e^{-\beta E}. \label{partial_inverse}
\end{equation}
The inverse Laplace transformation of the first
term is calculated by determining the singularities in terms of $\alpha$ and
than calculate their residues by using the Laurent expansion. Finding the poles
is here equivalent to determining the roots of the denominator,
\begin{equation}
g(\alpha,\beta)=(\alpha+\beta)\cosh(\sqrt{\alpha+\beta})=0.
\end{equation}
This is satisfied for 
\begin{equation}
\alpha=-\beta \qquad \text{and} \qquad \alpha_n=-\pi^2(n+\frac{1}{2})^2-\beta
\end{equation}
for $n \in \mathbb{N}$. And the residues are given by
\begin{eqnarray}
\text{Res}(\alpha g^{-1}(\alpha,\beta)e^{\alpha E},\alpha=-\beta) &=& -\beta e^{-\beta E}  \nonumber \\
\text{Res}(\alpha g^{-1}(\alpha,\beta)e^{\alpha E},\alpha_0=-\frac{\pi^2}{4}-\beta) &=& \frac{\pi^2+4\beta}{\pi}e^{-(\frac{\pi^2}{4}+\beta)E} \nonumber \\ 
\text{Res}(\alpha g^{-1}(\alpha,\beta)e^{\alpha E},\alpha_1=-\frac{-9\pi^2}{4}-4\beta)&=&-\frac{9\pi^2+4\beta}{5\pi}e^{-(\frac{9\pi^2}{4}+\beta)E}\\
& &\vdots \nonumber 
\end{eqnarray}
The first term above cancels with the last term in Eq.~\ref{partial_inverse}.
Combining the remaining terms, we are left with the overall enzyme exposure
distribution 
\begin{equation}
	P_{\rm u}(E)=\sum_{n=0}^{\infty} (-1)^n \bigg(    \frac{(2n+1)^2\pi^2+4\beta}{(2n+1)\pi}  \bigg) e^{-(\pi^2(n+\frac{1}{2})^2+\beta)E}. 
\end{equation}

\section{Three dimensions}

In three dimensions we impose rotational symmetry and reduce the reaction
diffusion equation to depend only on the radial coordinate, 
\begin{equation}
r^{-2}\partial_r (r^2 \partial_r \rho(r))-\alpha e(r)\rho(r)=0.
\label{appendix_3d_cluster}
\end{equation}
We apply the following boundary conditions, $(4\pi r^2 \partial_r
\rho(r))|_{r=0}=-1$ and the outer sphere is absorbing $\rho(r)|_{r=1}=0$.  

\subsection{Clustered enzyme profile with absorbing boundary}

We again begin by considering a clustered distribution of enzymes, 
$e_{\rm c}(r)=\frac{\delta(r-r_0)}{3r_0^2}$, that has been normalized such that
$\int_0^1 4\pi r^2 e_{\rm c}(r) {\rm d}r=4\pi/3$. As in section
\ref{clu_enzyeme_section} we divide the system into two parts, part $I$ for
$r<r_0$ and part $II$ for $r>r_0$. In each part, due to the absence of $E_2$
enzymes, the solution of Eq.~\ref{appendix_3d_cluster} is
\begin{equation}
\rho_i(r)=\frac{A_i}{r}+B_i
\end{equation}
where $i=\{ I,II \}$.
Applying the boundary conditions leads to the following two conditions 
\begin{equation}
	A_I=\frac{1}{4\pi} \qquad \text{and} \qquad A_{II}=-B_{II}. \label{3d_bcs}
\end{equation}
The remaining two conditions come again from the matching of the concentration
at $r=r_0$, $\rho_I(r_0)=\rho_{II}(r_0)$, and the discontinuity of the
derivative of the concentration $\rho(r)$ at $r=r_0$. Hence we get
\begin{equation}
\frac{1}{4\pi r_0}+B_I=A_{II}(\frac{1}{r_0}-1)
\end{equation} 
and
\begin{equation}
-A_{II}+\frac{1}{4
\pi}-A_{II}\left[\frac{\alpha}{3}\left(\frac{1}{r_0}-1\right)\right]=0.
\label{3d_deriv}
\end{equation}
With these four conditions Eqs.~\ref{3d_bcs}-\ref{3d_deriv} we obtain after some
straightforward algebra the expressions for the four constant
$\{A_{I},A_{II},B_{I},B_{II}\}$. Since we do not consider a competing pathway in
this model the efficiency of the pathway can also be calculated as 
\begin{equation}
	\bigg( \frac{J_2}{J_1} \bigg)_{\rm c}=1-4\pi (r^2\partial_r \rho(r))|_{r=1}=4\pi A_{II}=\frac{\frac{\alpha}{3}(1-r_0)}{r_0+\frac{\alpha}{3}(1-r_0)}.
\end{equation}

\subsection{Uniform  enzyme profile with absorbing boundary}

For the uniform enzyme profile $e_{\rm u}(r)=1$ the reaction-diffusion equation
Eq.~\ref{appendix_3d_cluster} reads
\begin{equation}
r^{-2}\partial_r( r^2 \partial_r \rho(r))-\alpha \rho(r)=0.
\end{equation}
This is solved by
\begin{equation}
\rho(r)=\frac{1}{r}\bigg( Ae^{\sqrt{\alpha}r}+Be^{-\sqrt{\alpha}r} \bigg).
\end{equation}
With the boundary conditions we arrive at the following two conditions: 
at the origin,
\begin{equation}
A+B=\frac{1}{4\pi},
\end{equation}
and at the absorbing outer boundary,
\begin{equation}
Ae^{\sqrt{\alpha}}+Be^{-\sqrt{\alpha}}=0.
\end{equation}
In the same way as the case of clustered enzymes the efficiency is given by
\begin{equation}
	\bigg( \frac{J_2}{J_1} \bigg)_{\rm u}=1-4\pi (r^2\partial_r
	\rho(r))|_{r=1}=1-\sqrt{\alpha} \rm{csch}(\sqrt{\alpha}).
\end{equation}

\section{Enzyme exposure probability distribution in 3D}

Similarly to the one dimensional case the enzyme exposure probability
distribution is obtained by the inverse Laplace transformation of the loss
current through the boundary. The loss current of the clustered enzyme profile
reads
\begin{equation}
\frac{J_{\rm{loss}}}{J_1}=\frac{r_0}{r_0+\frac{\alpha}{3}(1-r_0)}
\end{equation}
which has the general form discussed in Section \ref{enzyme_exp_1d_section}.
Hence the enzyme exposure distribution is
\begin{equation}
	P_{\rm c}(E)=\frac{3r_0}{1-r_0}e^{-\frac{3r_0}{1-r_0}E}.
\end{equation}
For the uniform enzyme profile, the calculation again proceeds in the same way
as described above.  We firstly calculate the singularities of the loss current,
the roots of $\sinh(\sqrt{\alpha})$, which are given by
\begin{equation}
\alpha_n=-(n\pi)^2 \qquad \text{with} \quad n \in \mathbb{N}.
\end{equation}
This leads then to the following residues
\begin{eqnarray}
\text{Res}(\frac{\sqrt{\alpha}}{\sinh\sqrt{\alpha}}e^{\alpha E},\alpha_0=0)&=&0 \nonumber \\
\text{Res}(\frac{\sqrt{\alpha}}{\sinh\sqrt{\alpha}}e^{\alpha E},\alpha_1=-\pi^2)&=& 2\pi^2e^{-\pi^2 E}\nonumber \\
\text{Res}(\frac{\sqrt{\alpha}}{\sinh\sqrt{\alpha}}e^{\alpha E},\alpha_2=-4\pi^2)&=& -8\pi^2e^{-4\pi^2 E}\\
\vdots\nonumber 
\end{eqnarray}
We assemble all the individual terms and get for the enzyme exposure probability distribution
\begin{equation}
	P_{\rm u}(E)=2\sum_{n=1}^{\infty} (-1)^{n+1}(\pi n)^2 e^{-(n\pi)^2E}.
\end{equation}

\end{document}